\documentclass[aps,prb,twocolumn,eqsecnum,superscriptaddress]{revtex4}%
\usepackage{amsmath}
\usepackage{amssymb}
\usepackage{amsfonts}
\usepackage{amsthm}
\usepackage{graphicx}%
\providecommand{\U}[1]{\protect\rule{.1in}{.1in}}

\begin{document}
\title{Direct evaluation of the temperature dependence of the rate constant based on
the quantum instanton approximation}
\author{Marcin Buchowiecki}
\affiliation{Laboratory of Theoretical Physical Chemistry, Institut des Sciences et
Ing\'{e}nierie Chimiques, \'{E}cole Polytechnique F\'{e}d\'{e}rale de Lausanne
(EPFL), CH-1015 Lausanne, Switzerland}
\affiliation{Institute of Physics, University of Szczecin, Wielkopolska 15, 70-451
Szczecin, Poland}
\author{Ji\v{r}\'{\i} Van{\'{\i}}{\v{c}}ek}
\email{jiri.vanicek@epfl.ch}
\affiliation{Laboratory of Theoretical Physical Chemistry, Institut des Sciences et
Ing\'{e}nierie Chimiques, \'{E}cole Polytechnique F\'{e}d\'{e}rale de Lausanne
(EPFL), CH-1015 Lausanne, Switzerland}
\date{\today}

\begin{abstract}
A general method for the direct evaluation of the temperature
dependence of the quantum-mechanical reaction rate constant in many-dimensional systems is described. The
method is based on the quantum instanton approximation for the rate constant,
thermodynamic integration with respect to the inverse temperature, and the
path integral Monte Carlo evaluation. It can describe deviations from the Arrhenius law
due to the coupling of rotations and vibrations, zero-point energy, tunneling,
corner-cutting, and other nuclear quantum effects. The method is tested on the Eckart barrier
and the full-dimensional H $+~$H$_{2}\rightarrow$~H$_{2}+$ H reaction. In the
temperature range from $300%
\operatorname{K}%
$ to $1500%
\operatorname{K}%
$, the error of the present method remains within $13\%$ despite the very large
deviations from the Arrhenius law. The direct approach makes the calculations
much more efficient, and the efficiency is increased even further (by up to two orders of magnitude in the studied reactions) by using optimal estimators for reactant and transition state thermal energies. Which of the estimators is optimal, however, depends on the system and the strength of constraint in a constrained simulation.

\end{abstract}
\maketitle


\section{Introduction}

The measurement of the temperature dependence of the rate constant is one of
the important tools of chemical kinetics in determining mechanisms of chemical
reactions.\cite{Elsamra2005,Persky2007,Baasandorj2009} Significant deviations from a simple exponential behavior can be
evidence of tunneling and of other nuclear quantum 
effects.\cite{Clary2008,Hu2006} These effects are particularly strong for
hydrogen transfer reactions with a high activation barrier or at low
temperatures. Recently, however, quantum effects have been observed also in many
enzymatic reactions at physiological
temperatures.\cite{Kohen1999,Basran1999,Marcus2006,Gao2008} It therefore
becomes more and more important to have accurate theoretical methods for
computing the temperature dependence of the rate constant.\cite{Ellingson2007}

Probably the oldest yet still the best known expression for the thermal rate
constant $k(T)$ at temperature $T$ is the empirical Arrhenius
law,\cite{Arrhenius1889}%
\begin{equation}
k_{A}(T)=Ae^{-E_{a}/k_{B}T}. \label{k_arrh}%
\end{equation}
Here $k_{B}$ is the Boltzmann constant, $E_{a}$ the activation energy, and 
the temperature dependence is purely exponential. An improvement over the
Arrhenius law was provided by the transition state theory
(TST),\cite{Eyring1935,Evans1935,Wigner1938} in which
\begin{equation}
k_{\text{TST}}(T)=\frac{k_{B}T}{h}\frac{Q^{\ddag}(T)}{Q_{r}(T)}e^{-\Delta
E^{\ddag}/k_{B}T}, \label{k_tst}%
\end{equation}
where $h$ is the Planck's constant, $Q^{\ddag}(T)$ and $Q_{r}(T)$ are the
partition functions of the transition state and the reactants, respectively,
and $\Delta E^{\ddag}$ is the barrier height for the reaction. Here the
temperature dependence includes a fractional power $T^{\alpha}$ in addition to 
the
exponential. Nevertheless, both Arrhenius law and TST are basically purely
classical, so they cannot take into account tunneling and other
nuclear quantum effects.

Although the simplest quantum effects \emph{can} be taken into account within
the TST in an \textit{ad hoc} fashion, by replacing the partition functions by
their quantum analogs for the simple harmonic oscillator (which takes into
account the zero-point energy, the Wigner tunneling
correction,\cite{Wigner1932} and approximate quantization of the
vibrational motion), a more systematic approach requires quantum treatment of
the nuclear motion. This is of course extremely difficult, and therefore
various approximate yet accurate methods have been developed. These include,
e.g., the semiclassical methods\cite{Miller1974,Chapman1975,Miller2001,Hu1994} or the so-called
quantum transition state
theories.\cite{Tromp1986,Voth1989a,Hwang1991,Truong1993,Hansen1996,Krilov2001,Wu2004,Schubert2009,Miller2003}%

In this paper, we evaluate the temperature dependence of the rate constant
starting from the quantum instanton (QI) approximation.\cite{Miller2003} This
quantum transition state theory has been shown to describe
correctly not only all of the above-mentioned quantum effects, but also
corner-cutting, coupling of vibrational and rotational motions, multiple
tunneling paths, etc. As we evaluate the temperature dependence of the rate
constant directly, we can speed up the QI\ calculation significantly by
avoiding the tedious umbrella sampling necessary for computing the rate constant itself.
Furthermore, if it is only the temperature dependence of the rate constant that is
needed, we can increase the accuracy of the QI approximation for the rate
constant by canceling some small remaining systematic errors. This can be
useful, e.g., if we know $k(T_{0})$ at a temperature $T_{0}$ very accurately
and would like to know $k(T)$ at other temperatures. The temperature dependence of the rate constant is computed via a thermodynamic integration \cite{Chandler1987,Frenkel2002} with respect to the inverse temperature. A similar thermodynamic integration in the framework of the QI model was used by Ceotto and Miller to compute the rate constant for several one and two-dimensional systems using a discrete variable representation.\cite{Ceotto2004} Below, we develop a general thermodynamic integration procedure based on the path integral implementation, which is suitable for many-dimensional systems.    

The method is tested on two simple systems for which exact quantum dynamical
calculations are feasible: the Eckart barrier and the full nine-dimensional H
$+~$H$_{2}\rightarrow$~H$_{2}+$ H reaction. While the former system is the
simplest one-dimensional model of a bimolecular reaction, the latter is the
simplest bimolecular chemical reaction with an energy barrier. As such, it has
been widely studied and attained the status of a benchmark
reaction.\cite{Mielke2003,Aoiz2005} Despite the apparent simplicity, this reaction remains
a challenging test for new approximations. This is due to the presence of strong
quantum effects, as the lightest atoms are involved in both bond breaking and
bond formation. This reaction was investigated not only for the temperature
dependence of its rate constant, but also for the kinetic isotope
effect,\cite{Yamamoto2004,Vanicek2005,Michael1990} the presence of the
geometric phase effect,\cite{Kendrick2003,Lepetit1990} etc.

The remainder of this paper is organized as follows: Section II describes the
methodology, i.e., the QI approximation, the thermodynamic integration with respect to the inverse temperature, the path integral
formalism, and the relevant estimators. Computational details and, in particular, the analysis of various errors are presented in Sec. III. Section IV contains the results.
First, it is shown how the low and high temperature limits are
obtained. The numerical results for the Eckart barrier and the
H$\,+\,$H$_{2}$ reaction are then presented. The results are compared with the
Arrhenius law, TST, TST with the Wigner tunneling correction, and the exact
quantum calculation. Finally, we discuss how the
efficiency is related to the dependence of the statistical error on the number
of imaginary time slices in the path integral, and how this error, in turn,
depends on the system under study. Section V concludes the paper.

\section{Methodology}

\subsection{\label{subsection:QI}Quantum instanton approximation for the
thermal rate constant}

The QI approximation for thermal rate constants was introduced
in Ref. \onlinecite{Miller2003}. The most direct derivation
\cite{Ceotto2005,Vanicek2005,M.Ceotto} starts from the exact Miller-Schwartz-Tromp
formula for the rate constant,\cite{Miller1983}%
\begin{equation}
k(T)Q_{r}=\int_{0}^{\infty}dt\,C_{\text{ff}}\left(\beta,  t\right)
\label{miller_schwartz_tromp}%
\end{equation}
where $C_{\text{ff}}\left(\beta,  t\right)  $ is the symmetrized flux-flux
correlation function,%
\begin{equation}
C_{\text{ff}}\left(\beta,  t\right)  =\operatorname{Tr}\left(  e^{-\beta\hat{H}%
/2}\hat{F}_{a}e^{-\beta\hat{H}/2}e^{i\hat{H}t/\hbar}\hat{F}_{b}e^{-i\hat
{H}t/\hbar}\right)  , \label{c_ff}%
\end{equation}
with Hamiltonian operator $\hat{H}$, inverse temperature $\beta:=1/k_{B}T$, time $t$,
and $\hat{F}_{\gamma}$ the flux operator through the dividing surface $\gamma
$. A stationary-phase approximation applied to Eq. (\ref{miller_schwartz_tromp}%
) yields the QI approximation for the rate constant,\cite{Vanicek2005,M.Ceotto}
\begin{equation}
k(T)\approx k_{\text{QI}}(T)=\frac{1}{Q_{r}}C_{\text{ff}}\left(\beta,  0\right)
\frac{\sqrt{\pi}}{2}\frac{\hbar}{\Delta H(\beta)} \label{k_QI}%
\end{equation}
where $\Delta H(\beta)$ is a specific type of energy variance,\cite{Yamamoto2004}
\begin{equation}
\Delta H (\beta)=\hbar\left[  \frac{-\ddot{C}_{dd}\left(\beta,  0\right)  }{2C_{dd}\left(\beta,
0\right)  }\right]  ^{1/2}. \label{dh}%
\end{equation}
The delta-delta correlation function used above is defined as
\begin{equation}
C_{\text{dd}}\left(\beta,  t\right)  =\operatorname{Tr}\left(  e^{-\beta\hat{H}%
/2}\hat{\Delta}_{a}e^{-\beta\hat{H}/2}e^{i\hat{H}t/\hbar}\hat{\Delta}%
_{b}e^{-i\hat{H}t/\hbar}\right)  \label{c_dd}%
\end{equation}
where the generalized delta function operator is given by
\begin{equation}
\hat{\Delta}_{\gamma}=\Delta\left[  \xi\left(  \mathbf{\hat{r}}\right)
-\xi_{\gamma}\right]  \equiv\delta\left[  \xi\left(  \mathbf{\hat{r}}\right)
-\xi_{\gamma}\right]  m^{-1/2}\left\Vert \nabla\xi\right\Vert \label{delta_op}%
\end{equation}
and $\xi(\mathbf{r})$ is the reaction coordinate such that $\xi(\mathbf{r}%
^{\ddag})=0$ at the transition state. Similarly, $\xi(\mathbf{r})=\xi_{\gamma
}$ defines the position of the dividing surface $\gamma$. We have used
mass-scaled coordinates in which all degrees of freedom have the same mass
$m$. In practice, the exact delta function constraint is approximated by a
Gaussian constraint corresponding to a harmonic constraint potential
$V_{\text{constr}}(\mathbf{r})$,\cite{Vanicek2005,Yamamoto2004}%
\begin{align}
\delta\lbrack\xi(\mathbf{r})-\xi_{\gamma}]  &  \approx 
\sqrt{\frac{\beta}{2\pi \sigma^2}}
e^{-\beta V_{\text{constr}}(\mathbf{r})},\label{gauss}\\
V_{\text{constr}}(\mathbf{r})  &  =\frac{1}{2}\left(  \frac{\xi(\mathbf{r}%
)-\xi_{\gamma}}{\sigma}\right)  ^{2}. \label{v_constr}%
\end{align}

The accuracy of the QI approximation has been already verified
in numerous
applications.\cite{Miller2003,Yamamoto2004,Zhao2004,Yamamoto2005,Vanicek2005,Li2004,Vanicek2007,Wang2009,Zimmermann2010}
The main shortcoming of the QI method is the neglect of recrossing which is
however, neglected in any quantum or classical transition state theories. The recrossing effects on the quantum instanton rate constant have been quantified for several collinear reactions by Ceotto and Miller.\cite{Ceotto2004} Fortunately, the recrossing effects become generally less important in higher dimensions.

\subsection{\label{subsection:TI}Temperature dependence via the thermodynamic
integration}

The goal of this paper is to compute the temperature dependence of the rate
constant, i.e., the ratio $k(T_{2})/k(T_{1})$. Within the QI approximation,
this ratio is given by%

\begin{equation}
\frac{k(T_{2})}{k(T_{1})}=\frac{Q_{r}(\beta_{1})}{Q_{r}(\beta_{2}%
)}\frac{\Delta H(\beta_{1})}{\Delta H(\beta_{2})}\frac{C_{\text{dd}}\left(
\beta_{2},0\right)  }{C_{\text{dd}}\left(  \beta_{1},0\right)  }\frac
{\frac{C_{\text{ff}}\left(  \beta_{2},0\right)  }{C_{\text{dd}}\left(  \beta
_{2},0\right)  }}{\frac{C_{\text{ff}}\left(  \beta_{1},0\right)  }{C_{\text{dd}%
}\left(  \beta_{1},0\right)  }}, \label{qi_tdep}%
\end{equation}
where we multiplied and divided the numerator and denominator by
$C_{\text{dd}}(\beta,0)$. In this expression, quantities $\Delta H(\beta)$ and $C_{\text{ff}}\left(
\beta,0\right)  /C_{\text{dd}}\left(  \beta,0\right)  $ can be computed directly
by the Metropolis Monte-Carlo procedure because they are thermodynamic averages. On the other hand,
the ratios $Q_{r}(\beta_{1})/Q_{r}(\beta_{2})$ and $C_{\text{dd}}\left(
\beta_{2},0\right)  /C_{\text{dd}}\left(  \beta_{1},0\right)  $ cannot be computed this way since they
involve ratios of quantities at different temperatures. These ratios can,
however, be calculated by the method of \emph{thermodynamic integration}%
~\cite{Frenkel2002,Chandler1987} with respect to the inverse temperature
$\beta$,%
\begin{align}
\frac{Q_{r}(\beta_{2})}{Q_{r}(\beta_{1})}  &  =\exp\left[  -\int_{\beta_{1}%
}^{\beta_{2}}E_{r}(\beta)d\beta\right]  ,\label{ti_qr}\\
\frac{C_{\text{dd}}\left(  \beta_{2},0\right)  }{C_{\text{dd}}\left(  \beta
_{1},0\right)  }  &  =\exp\left[  -\int_{\beta_{1}}^{\beta_{2}}E^{\ddag}%
(\beta)d\beta\right]  , \label{ti_cdd}%
\end{align}
where $E_{r}$ and $E^{\ddag}$ are the thermal energies of the reactants and
of the transition state, respectively. A similar thermodynamic integration was used within a discrete variable representation of the QI approximation to compute the rate constant for several collinear triatomic reactions.\cite{Ceotto2004} Unlike $Q_{r}$ and $C_{\text{dd}}$, the energies are normalized quantities because they can be written as logarithmic
derivatives:%
\begin{align}
E_{r}(\beta)  &  :=-\frac{d\log Q_{r}(\beta)}{d\beta}=-\frac{dQ_{r}%
(\beta)/d\beta}{Q_{r}(\beta)},\label{er_logder}\\
E^{\ddag}(\beta)  &  :=-\frac{d\log C_{\text{dd}}(\beta,0)}{d\beta}%
=-\frac{dC_{\text{dd}}(\beta,0)/d\beta}{C_{\text{dd}}(\beta,0)}.
\label{ets_logder}%
\end{align}
Hence they can be computed directly by a Monte Carlo procedure.

\subsection{\label{subsection:PIMC}Path integral representation of relevant
quantities}

Quantum thermodynamic effects can be treated rigorously using the imaginary
time path integral (PI).\cite{Feynman1965,Topper1999,Berne1986,Kleinert2004,Ceperley1995}
Let $D$ be the number of degrees of freedom ($D=1$ for the Eckart barrier and
$D=9$ for the H$_{3}$ potential) and $P$ the number of imaginary time slices
in the PI. The PI representations of the partition
function\cite{Topper1999,Berne1986,Kleinert2004,Ceperley1995} and the
delta-delta correlation function\cite{Yamamoto2004} are%
\begin{align}
Q_{r}^{P}(\beta)  &  =C\int d\mathbf{r}^{\left(  1\right)  }\cdots\int d\mathbf{r}%
^{\left(  P\right)  }\exp\left[  -\beta\Phi\left(  \left\{  \mathbf{r}%
^{\left(  s\right)  }\right\}  \right)  \right]  ,\label{pi_qr}\\
C_{\text{dd}}^{P}\left(\beta,  0\right)   &  =C\int d\mathbf{r}^{\left(  1\right)
}\cdots\int d\mathbf{r}^{\left(  P\right)  }\Delta\left[  \xi_{a}\left(
\mathbf{r}^{\left(  0\right)  }\right)  \right]  \nonumber \\  & \times \Delta\left[  \xi_{b}\left(
\mathbf{r}^{\left(  P/2\right)  }\right)  \right]  \exp\left[  -\beta
\Phi\left(  \left\{  \mathbf{r}^{\left(  s\right)  }\right\}  \right)
\right]  , \label{pi_cdd}%
\end{align}
where $C=[mP/(2\pi\hbar^{2}\beta)]^{DP/2}$, $\mathbf{r}^{\left(  s\right)  }$
is a $D$-dimensional vector representing the $s$th time slice, and the
effective potential $\Phi$ is given by%
\begin{equation}
\Phi\left(  \left\{  \mathbf{r}^{\left(  s\right)  }\right\}  \right)
=\frac{mP}{2\hbar^{2}\beta^{2}}\sum_{s=1}^{P}\left(  \mathbf{r}^{\left(
s\right)  }-\mathbf{r}^{\left(  s-1\right)  }\right)  ^{2}+\frac{1}{P}%
\sum_{s=1}^{P}V\left(  \mathbf{r}^{\left(  s\right)  }\right)  . \label{phi}%
\end{equation}
For $P=1$, the above expressions reproduce classical statistical mechanics,
while exact quantum statistics is reached in the limit $P\rightarrow\infty$.

In practice, there are two main strategies for evaluating thermodynamic
averages using the PI: the PI\ molecular dynamics (PIMD)\cite{Berne1986,Tuckerman2000} or PI Monte Carlo
(PIMC).\cite{Ceperley1995} We use the PIMC procedure together with the Metropolis algorithm. The
basic idea is to sample the PI\ configuration space according to an
appropriate weight $\rho$, which is, e.g., for $C_{\text{dd}}$ given by
\begin{equation}
\rho^{\ddag}\left(  \left\{  \mathbf{r}^{\left(  s\right)  }\right\}  \right)
:=\Delta\left[  \xi_{a}\left(  \mathbf{r}^{\left(  0\right)  }\right)
\right]  \Delta\left[  \xi_{b}\left(  \mathbf{r}^{\left(  P/2\right)
}\right)  \right]  \exp\left[  -\beta\Phi\left(  \left\{  \mathbf{r}^{\left(
s\right)  }\right\}  \right)  \right]  , \label{rho_ts}%
\end{equation}
and then, at each sampled configuration, to evaluate the so-called estimator
$A^{P}(\left\{  \mathbf{r}^{(s)}\right\}  )$ of the relevant physical quantity
$A$. The final estimate of $A$ is given by the average $\left\langle
A^{P}\right\rangle $ along the PIMC trajectory.

Using the PI representation of $Q_{r}$ and $C_{\text{dd}}$, one can
obtain\ estimators for all quantities needed in Eq. (\ref{qi_tdep}), i.e.,
$\Delta H$, $C_{\text{ff}}/C_{\text{dd}}$, and the logarithmic derivatives of
$Q_{\text{r}}$, $C_{\text{dd}}$ (i.e., the energies $E_{r}$, $E^{\ddag}$).
Those for $\Delta H$ and $C_{\text{ff}}/C_{\text{dd}}$ are listed in
Ref. \onlinecite{Yamamoto2004}.

\subsection{Estimators for $E_{r}$}

The simplest estimator for the energy $E_{r}$, the so-called Barker or thermodynamic
estimator (TE),\cite{Barker1979} can be derived directly from Eq.
(\ref{er_logder}) and the PI expression (\ref{pi_qr}), giving%
\begin{equation}
E_{r,\text{TE}}^{P}=\frac{DP}{2\beta}-\frac{mP}{2\beta^{2}}\sum_{s=1}%
^{P}(\mathbf{r}^{(s)}-\mathbf{r}^{(s-1)})^{2}+\frac{1}{P}\sum_{s=1}%
^{P}V(\mathbf{r}^{(s)}). \label{Eth}%
\end{equation}
As observed by Herman \textit{et al.},\cite{Herman1982} the TE can have a
large statistical error, which can be avoided with the so-called virial
(VE)\cite{Herman1982} or centroid virial (CVE),\cite{Parrinello1984} estimators.  
Invoking the virial theorem,  the kinetic energy in these two
estimators is replaced by an expression involving the
gradient of the potential energy.\cite{Herman1982} 
This is convenient in the
PIMD implementations since the gradient is already available. In PIMC
simulations, however, only the potential is needed for the random walk, and in
order to avoid computing the gradients, alternative approaches have been proposed. One can, e.g., employ the centroid thermodynamic estimator\cite{Glaesemann2002} or more generally, use a procedure based on rescaling coordinates\cite{Predescu2002,Predescu2003a} in which the gradients of the potential are replaced by a single derivative that can be evaluated by finite difference.\cite{Predescu2003b} Variants of the latter approach have been applied successfully to compute thermal energies and heat
capacities,\cite{Predescu2003b} kinetic isotope
effects\cite{Vanicek2005a,Vanicek2007}, equilibrium isotope
effects\cite{Zimmermann2009}, or the derivatives of the flux-flux correlation function\cite{Yang2006} needed in the generalized QI model.\cite{Ceotto2005}

The VE for $E_{r}$ can be derived most directly by the change of coordinates
$\mathbf{x}^{(s)}:=\beta^{-1/2}\mathbf{r}^{(s)}$ in the PI (\ref{pi_qr}),
yielding%
\begin{align}
E_{r,\text{VE}}^{P} & = \frac{1}{P} \sum_{s=1}^{P}
\left\{ 
V(\mathbf{r}^{(s)}) +
\beta \frac{dV[(\beta+\Delta \beta)^{1/2}\beta^{-1/2}\mathbf{r}^{(s)}]}{d\Delta\beta} 
\right\} \nonumber \\
 & = \frac{1}{P} \sum_{s=1}^{P}
\left\{
V(\mathbf{r}^{(s)}) +
\frac{dV[(1+q)^{1/2}\mathbf{r}^{(s)}]}{dq}
\right\}, 
\label{Ev}%
\end{align}
where $q$ is a small dimensionless parameter and the $q$-derivative is evaluated by finite difference at $q=0$.
Similarly, the CVE can be obtained by the change of variables $\mathbf{x}%
^{(s)}:=\beta^{-1/2}(\mathbf{r}^{(s)}-\mathbf{r}^{(C)})$, where one first
subtracts the so-called centroid coordinate $\mathbf{r}^{(C)}:=P^{-1}%
\sum_{s=1}^{P}\mathbf{r}^{(s)}$. The resulting estimator is%
\begin{align}
&E_{r,\text{CVE}}^{P}=\frac{D}{2\beta} +
\label{Ecv}\\
& + \frac{1}{P}\sum_{s=1}^{P}
\left\{
V(\mathbf{r}^{(s)}) +
\frac
{dV[\mathbf{r}^{(C)}+(1+q)^{1/2}(\mathbf{r}%
^{(s)}-\mathbf{r}^{(C)})]}{dq}
\right\}. \nonumber
\end{align}

\subsection{Estimators for $E^{\ddag}$}

In the case of constrained simulations near the transition state, the constrained weight
function (\ref{rho_ts}) can be approximated by using the Gaussian
approximation of the delta function from Eqs. (\ref{gauss})-(\ref{v_constr}).
Besides a prefactor, this amounts to adding a constraint potential%
\[
\Phi_{\text{constr}}(\{\mathbf{r}^{(s)}\}):=V_{\text{constr}}(\mathbf{r}%
^{(P/2)})+V_{\text{constr}}(\mathbf{r}^{(P)})
\]
to the effective potential $\Phi$. Assuming that $V_{\text{constr}}$ is
independent of temperature and following a derivation similar to that for
estimators of $E_{r}$, one obtains the thermodynamic, virial, and centroid
virial estimators for $E^{\ddag},$%
\begin{align}
E_{\text{TE}}^{\ddag,P}  &  =E_{r,\text{TE}}^{P}-\frac{1}{\beta}%
+\Phi_{\text{constr}}(\{\mathbf{r}^{(s)}\}),\label{e_TS_TE_rig}\\
E_{\text{VE}}^{\ddag,P}  &  =E_{r,\text{VE}}^{P}-\frac{1}{\beta}%
+\Phi_{\text{constr}}(\{\mathbf{r}^{(s)}\}) \nonumber \\ &  +
\frac{d\Phi_{\text{constr}%
}[(1+q)^{1/2}\{\mathbf{r}^{(s)}\}]}
{dq},\label{e_TS_VE_rig}\\
E_{\text{CVE}}^{\ddag,P}  &  =E_{r,\text{CVE}}^{P}-\frac{1}{\beta}%
+\Phi_{\text{constr}}(\{\mathbf{r}^{(s)}\}) \nonumber \\  &  +
\frac{d\Phi_{\text{constr}%
}[\{\mathbf{r}^{(C)}+(1+q)^{1/2}(\mathbf{r}%
^{(s)}-\mathbf{r}^{(C)})\}]}{dq}.
\label{e_TS_CVE_rig}%
\end{align}

Although the above estimators converge to the exact results, we found that the
statistical errors can be decreased slightly by employing an alternative set of
estimators, derived using an exact relation%
\[
\left\langle \Phi_{\text{constr}}(\{\mathbf{r}^{(s)}\})\right\rangle
=\beta^{-1},
\]
which is valid for a harmonic constraint potential for any value of $P$. The
new estimators are%
\begin{align}
E_{\text{TE}}^{\ddag,P}  &  =E_{r,\text{TE}}^{P},\label{e_TS_TE_adh}\\
E_{\text{VE}}^{\ddag,P}  &  =E_{r,\text{VE}}^{P}+ \frac{d\Phi_{\text{constr}%
}[(1+q)^{1/2}\{\mathbf{r}^{(s)}\}]}{dq},
\label{e_TS_VE_adh}\\
E_{\text{CVE}}^{\ddag,P}  &  =E_{r,\text{CVE}}^{P}+ \frac{d\Phi
_{\text{constr}}[\{\mathbf{r}^{(C)}+(1+q)^{1/2}%
(\mathbf{r}^{(s)}-\mathbf{r}^{(C)})\}]}{dq}.
\label{e_TS_CVE_adh}%
\end{align}
Estimators (\ref{e_TS_TE_adh})-(\ref{e_TS_CVE_adh}) are in a way more
intuitive than estimators (\ref{e_TS_TE_rig})-(\ref{e_TS_CVE_rig}): in the
limit of a sharp constraint, the constrained energy should be independent of
the type of constraint.

It should be stressed that the last terms in the VE and CVE in Eqs.
(\ref{e_TS_VE_rig}), (\ref{e_TS_CVE_rig}), (\ref{e_TS_VE_adh}), and 
(\ref{e_TS_CVE_adh}) are important; without them the agreement among the
TE, VE, and CVE is lost. In other words, an intuitive guess such as 
$E_{\text{CVE}}^{\ddag,P}  = E_{r,\text{CVE}}^{P}$ would not give a correct answer 
for the constrained energy.
 
Finally, we also tested a constraint potential that is proportional to
temperature, i.e.,%
\begin{equation}
V_{\text{constr}}=\beta^{-1}\widetilde{V}_{\text{constr}},
\label{v_constr_propto_T}%
\end{equation}
where $\widetilde{V}_{\text{constr}}$ is a harmonic potential independent of
temperature. As a result, the constraint (\ref{gauss}) itself is actually
independent of temperature. Following again a derivation similar to that for
estimators of $E_{r}$, one obtains another set of the TE, VE, and CVE for
$E^{\ddag},$ that look exactly like the estimators (\ref{e_TS_TE_adh}%
)-(\ref{e_TS_CVE_adh}) for $V_{\text{constr}}$ independent of temperature. The
only difference is that the random walk is done with a different constraint.


\section{Computational details and error analysis}


All calculations were performed with a PIMC code implemented in Fortran 90.
Sampling of the configurational space in the PIMC simulation was done using
three types of moves. Staging algorithm \cite{Sprik1985} was employed to move
all unconstrained beads. Constrained beads, i.e., beads $s=P/2$ and $s=P$
which feel the constraint potential $V_{\text{constr}}$, were sampled with the
free particle single slice algorithm.\cite{Ceperley1995} Finally, whole chain
moves\cite{Ceperley1995} were used to speed up sampling of the potential
energy surface.

The Gaussian constraint potential must be strong enough in order to exert the
constraining effect on the system. When this condition was satisfied, the
converged results were independent of the constraint. However, the statistical root mean square error (RMSE) of the
transition state energy $E^{\ddag}$ increases with the strength of the
constraint because sampling of the configuration space becomes more difficult.
Therefore the selected strength of the constraint should take into account
these two effects. We have used $k=10$ a.u. in both systems.

All quantities needed in the ratio (\ref{qi_tdep}) were evaluated using the
above mentioned estimators. The thermodynamic integrations (\ref{ti_qr}) and
(\ref{ti_cdd}) were evaluated with the Simpson rule using 25 values of $\beta$
between $\beta_{0}=1/k_{B}T_{0}$ and $\beta_{\text{max}}=1/k_{B}T_{\text{min}%
}$ with the reference temperature $T_{0}=1500%
\operatorname{K}%
$ and the minimum temperature $T_{\text{min}}=200%
\operatorname{K}%
$. The number of beads was chosen to be inversely proportional to the
temperature, with the maximal number of beads (used for $T_{\text{min}}=200%
\operatorname{K}%
$) being $P=96$ for the Eckart barrier and $P=160$ for the H$\,+\,$H$_{2}$ reaction.

The error of the final result consists of four main error contributions: a)
the statistical error due to the Monte Carlo simulation, b) the error due to
the discretization of the TI, c) the error due to the discretization of the PI
(i.e., the \textquotedblleft finite $P$ error\textquotedblright), and d) the
actual error of the QI approximation. We have carefully separated these four
contributions and attempted to make the first three contributions small in
comparison with the error of the QI. In more complicated systems, this may not
be possible and especially the final statistical error may be comparable to or
larger than the error of the QI. Because the exponentiation of the TI is quite
sensitive to various errors, a detailed analysis of errors was carried out for
the ratio $k(200%
\operatorname{K}%
)/k(1500%
\operatorname{K}%
)$, i.e., over the largest temperature range, where the first three types of errors are the greatest. The TI was evaluated
by four different numerical methods, namely the trapezoidal, Simpson, Simpson
$3/8$, and Boole methods.\cite{Kress1998}

Comparing the analytical bounds on the \emph{discretization errors} of the
TI integrals using a numerical estimate of a higher order derivative,\cite{Kress1998} one can
conclude that both the Simpson and Simpson 3/8 methods were much better than the
trapezoidal rule and that the Boole method did not provide any further
improvement. Specifically, for the Eckart barrier the error of the ratio due to the discretization 
of the TI was $2\%$, $0.03\%$, $0.02\%$, or $0.04\%$ for the trapezoidal, Simpson, Simpson
$3/8$, or Boole methods, respectively. For the H\thinspace+\thinspace
H$_{2}$ reaction, the discretization error of the final ratio was $7\%$, $0.3\%$, $0.1\%$, or $0.3\%$, in the same order. It should be emphasized that these error estimates are very conservative, as the actual difference between the final ratios based on different methods was an order of magnitude smaller than what one would expect from the error estimates. The final results displayed in the plots used the Simpson method.

The \emph{statistical} RMSEs were estimated with the block
averaging method using a variable block size \cite{Flyvbjerg1989} to remove
correlation of the PIMC data. The statistical error of the TI was evaluated
using an appropriate formula for each integration method and assuming that the
statistical errors of energies at different temperatures were uncorrelated. As
expected, the statistical error did not depend much on the integration method,
and was always close to the statistical error for the Simpson method. The
statistical error of the final ratio was $0.3\%$ for the Eckart barrier and
$1.6\%$ for the H\thinspace+\thinspace H$_{2}$ reaction.

The \emph{finite }$P$\emph{ error}, i.e., the error due to the discretization
of the Feynman PI, was obtained by repeating calculations of all quantities at
all temperatures with twice smaller numbers of beads ($P\rightarrow P/2$) and
then extrapolating each quantity to $P\rightarrow\infty$, assuming $1/P^{2}$
convergence. We emphasize that we used the extrapolated results only for
estimating the finite $P$ error of the computed ratio and \emph{not }for
estimating the ratio itself, which could be dangerous. The finite $P$ error of
the ratio was $-0.3\%$ for the Eckart barrier and $-3.5\%$ for the
H\thinspace+\thinspace H$_{2}$ reaction.

[We note that for H\thinspace+\thinspace H$_{2}$ one of the temperatures
($972.93%
\operatorname{K}%
$, a temperature in the vicinity of which a sharp bend in the $E^{\ddag}%
-E_{r}$ dependence occurs) required a five times longer simulation to reduce
the TI discretization errors. This was because a small statistical error had a
huge effect on the estimate of the fourth derivative and hence on the
analytical estimate of the discretization error.]

To sum up, in both systems the TI discretization error was negligible to the
statistical and finite $P$ errors, which, in turn, were small in comparison to
the error of the QI approximation.


\section{Results}


\subsection{Temperature dependence according to the Arrhenius law, TST, and the TST with the Wigner tunneling correction}

At high temperatures, the rate constant is expected to behave classically and
follow the Arrhenius law or the more accurate TST result. Whereas the
Arrhenius law (\ref{k_arrh}) predicts the rate constant ratio to be
a simple exponential function of the inverse temperature,%

\begin{equation}
\frac{k_{A}(\beta_{2})}{k_{A}(\beta_{1})}
= e^{-E_{a}(\beta_{2}-\beta_{1})}, \label{tdep_arrh}%
\end{equation}
TST (\ref{k_tst}) gives the ratio of rate constants
\begin{equation}
\frac{k_{\text{TST}}(\beta_{2})}{k_{\text{TST}}(\beta_{1})}=\frac{\beta_{1}%
}{\beta_{2}}\frac{Q^{\ddag}(\beta_{2})}{Q^{\ddag}(\beta_{1})}\frac{Q_{r}%
(\beta_{1})}{Q_{r}(\beta_{2})}e^{-(\beta_{2}-\beta_{1})\Delta E^{\ddag}}.
\label{tdep_tst}%
\end{equation}
In particular, assuming the partition functions $Q^{\ddagger}$ and $Q_{r}$ to
be separable into products of classical rotational and vibrational partition
functions, the temperature dependence (\ref{tdep_tst})\ of TST rate constant
includes an additional fractional power law besides the exponential dependence
in the Arrhenius law (\ref{tdep_arrh}).

At somewhat lower temperatures, when quantum effects start to play a role, the
basic TST expression (\ref{tdep_tst}) can be improved in several ways: First,
classical partition functions $Q_{r}$ and $Q^{\ddag}$\ can be replaced by
their exact quantum analogs for a harmonic potential. Second, quantum
tunneling can be included approximately via the Wigner tunneling
correction.\cite{Wigner1932} This method corrects the rate constant with a
multiplicative factor%
\begin{equation}
\kappa=1+\frac{ h^2 |\nu^{\ddag}|^2 \beta^{2}}{24}, \label{kappa_wigner}%
\end{equation}
where $\nu^{\ddag}$ is the imaginary frequency of the asymmetric stretch along
the reaction coordinate. The correction can be derived by treating the motion
through the transition state as a vibration on an upside down potential and expanding the
quantum partition function to second order in $\beta$. Although an improvement
over TST, the Wigner tunneling correction cannot describe multidimensional tunneling.


\subsection{Eckart barrier}

A simple model of an activated chemical reaction is provided by the Eckart
barrier, a one-dimensional system described by the potential%

\begin{equation}
V(x)=V_{0}\left[  \operatorname{cosh}(ax)\right]  ^{-2}.
\end{equation}
We use standard parameter values $V_{0}=1.56\cdot10^{-2}\ \mathrm{a.u.}$,
$a=1.36\ \mathrm{a.u.}$, mass $m=1060\ \mathrm{a.u.}$ and reaction coordinate
$\xi:=x$. The exact quantum rate constant $k_{\text{QM}}$ for this reaction
can be obtained by integrating the exact quantum mechanical cumulative
reaction probability, which is known analytically.\cite{Johnston1966}

\begin{figure}[hptb]
\includegraphics[width=1\columnwidth]{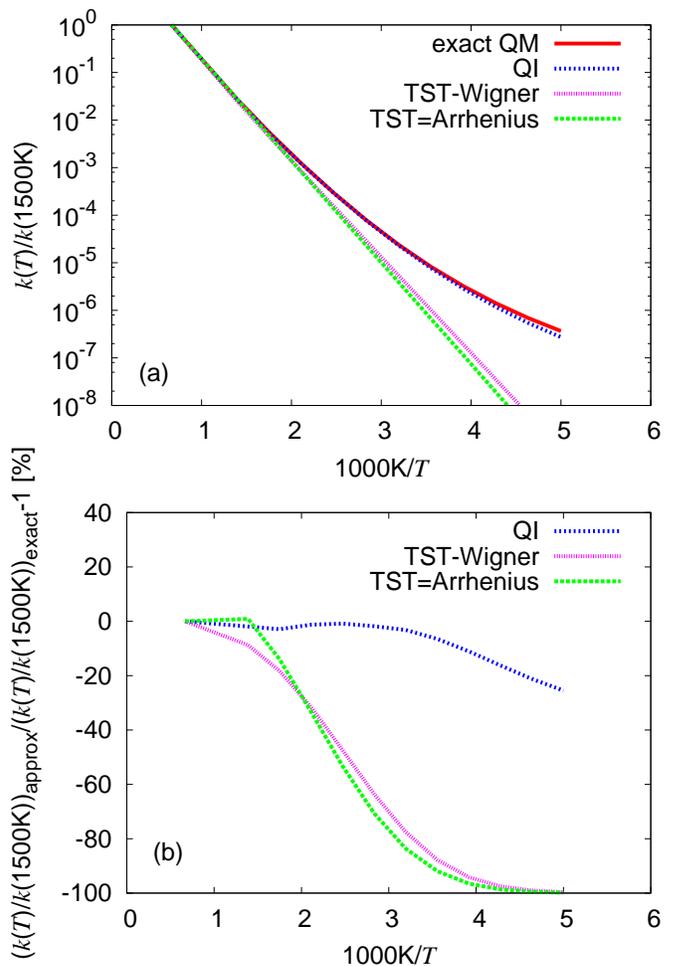} \caption{Eckart barrier. (a)
Temperature dependence of the rate constant. (b) Temperature dependence of the
relative error of the ratio $k(T)/k(1500\operatorname{K})$.}
\label{fig:eckart}%
\end{figure}

Figure \ref{fig:eckart} (a) compares the QI results with the exact QM results,
TST (which is equal to the Arrhenius law here), and the TST including the
Wigner tunneling correction. The reference temperature is $1500\,\mathrm{K}$ and
the plot shows ratios for temperatures down to $200\,\mathrm{K}$. Since
classical recrossing does not occur for the Eckart barrier, all TSTs should converge to the 
correct quantum results at high
temperatures. The
figure confirms that this is indeed the case:\ note that all curves are
tangent at the high temperature limit. At low temperatures, one reaches the
quantum regime where tunneling is important and consequently the Arrhenius
plot of the exact QM result has a large curvature. While TST has a huge error,
the QI approximation agrees very well with the QM result. Note that the Wigner
tunneling correction improves over the TST and captures the tunneling effect
partially but still fails to recover the curvature of the exact result.

Figure \ref{fig:eckart} (b) shows the relative error of the rate constant
ratio for the different methods. Whereas both TST and TST with the Wigner tunneling 
correction 
deteriorate rapidly with decreasing temperature, the QI method has an error
below $3\%$ for all temperatures above $330%
\operatorname{K}%
$. The QI approximation has a significant error ($\geq 10 \%$) only at very low temperatures, below 
$\sim 270\,\mathrm{K}$. However, this error was well understood already in the
original paper by Miller et al. \cite{Miller2003} and can be remedied by
considering two separate dividing surfaces at very low temperatures. (Here we have
used a single dividing surface at all temperatures for simplicity.)

\begin{figure}[hptb]
\includegraphics[width=1\columnwidth]{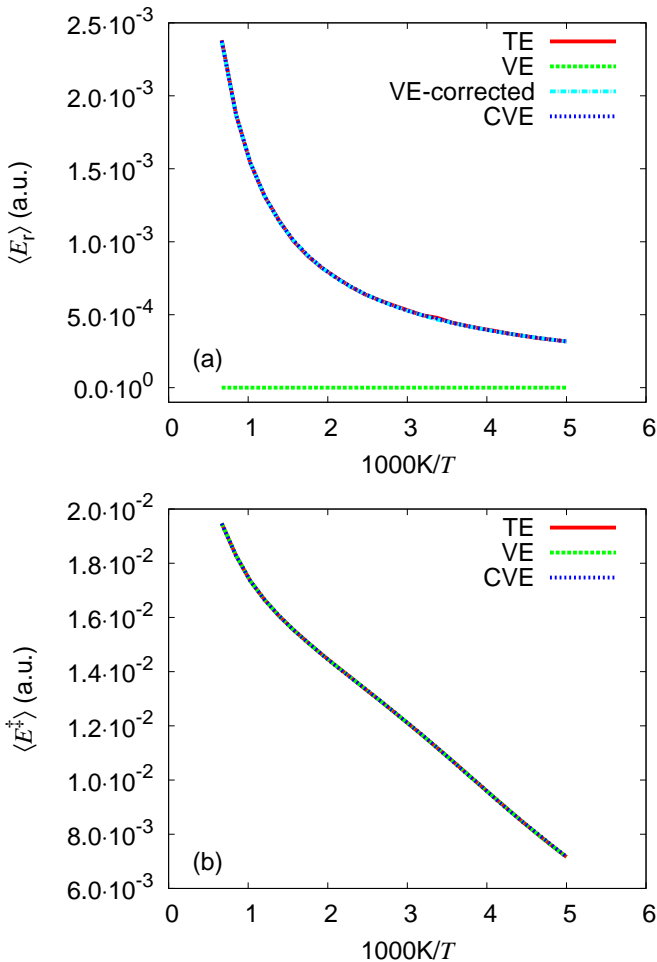} \caption{Eckart barrier. Temperature dependence of the reactant (a) and transition state (b) energies.}
\label{fig:eckart_ti}%
\end{figure}

The temperature dependence of the reactant and transition state energies is shown in Fig.~\ref{fig:eckart_ti}. While both curves are quite smooth, small discretization errors in the integrals can have large effects on the exponentiated result. By a detailed error analysis described in Sec. III, we found that the Simpson method was sufficient for the TI over the whole temperature range. Note that the VE for $E_r$ gives zero, but can be easily corrected with an analytical correction $1/(2 \beta)$.

\begin{figure}[hptb]
\includegraphics[width=1\columnwidth]{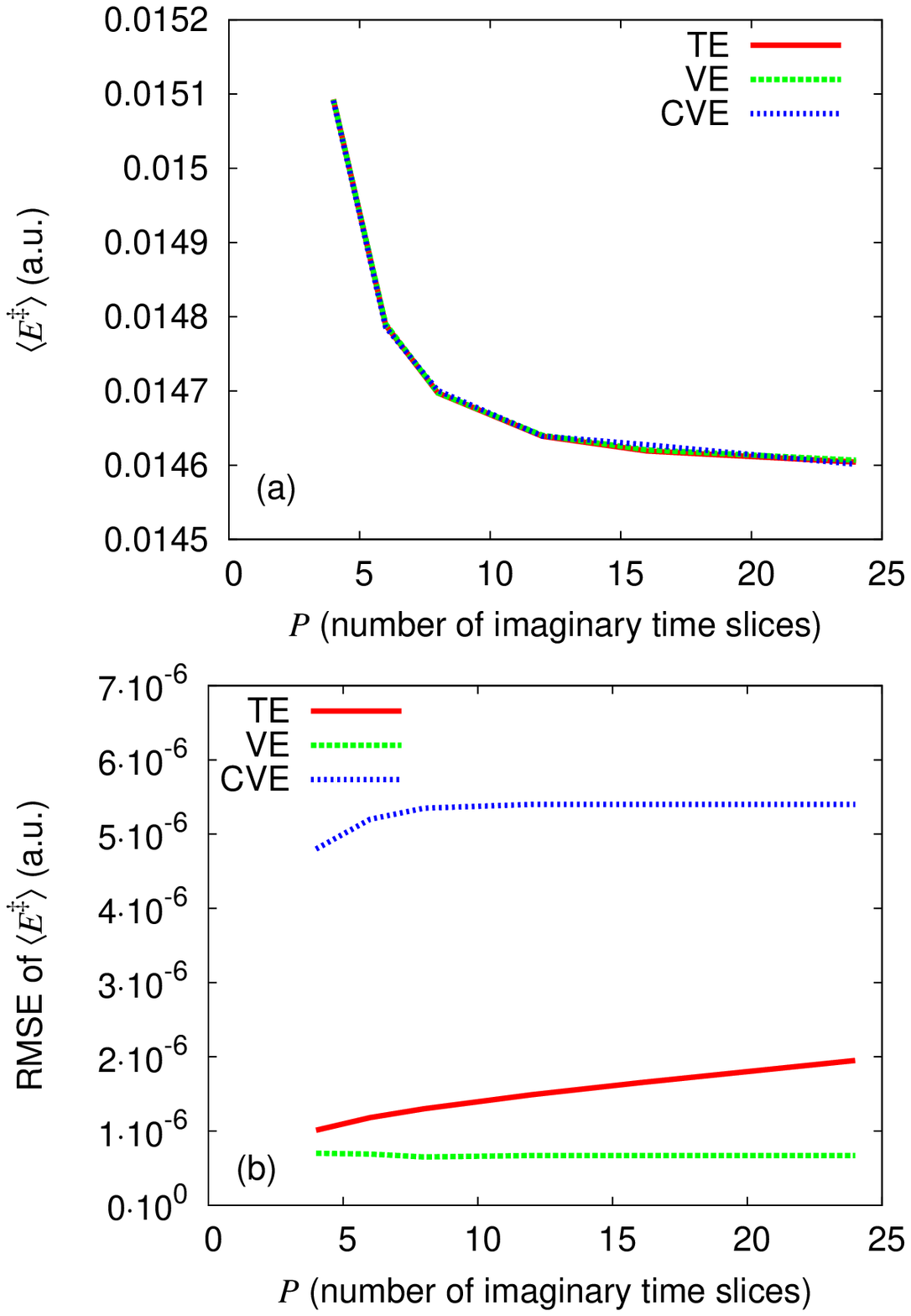} \caption{Dependence of the
transition state energy (a) and of the statistical RMSE of the transition state energy (b)
on the number of imaginary time slices for the Eckart barrier at
$T=515.15\operatorname{K}$. The constraint potential is independent of
temperature and estimators (\ref{e_TS_TE_adh})-(\ref{e_TS_CVE_adh}) are used.}
\label{fig:eckart_EvsP_adhoc}%
\end{figure}

\begin{figure}[ptb]
\includegraphics[width=1\columnwidth]{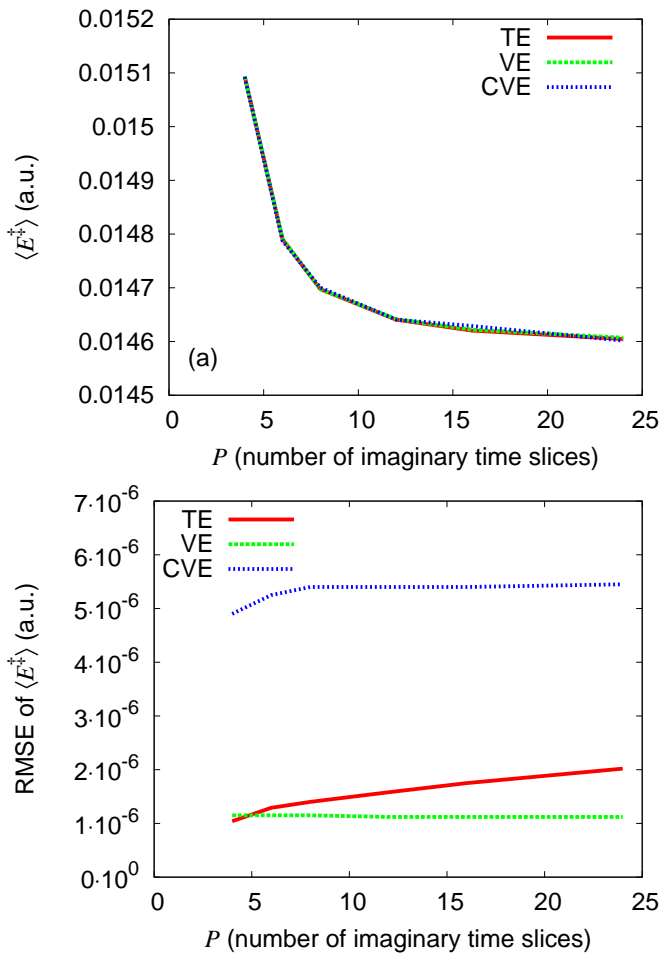} \caption{M. Buchowiecki and J. Van\'{\i}\v{c}ek}%
\label{fig:eckart_EvsP_rigorous}%
\end{figure}

The three different estimators for the constrained energy $E^{\ddag}$ at
$T=515.15\operatorname{K}$ are compared in Figs. \ref{fig:eckart_EvsP_adhoc}
and \ref{fig:eckart_EvsP_rigorous}. Panel (a) of Fig.
\ref{fig:eckart_EvsP_adhoc}, which uses the simpler estimators
(\ref{e_TS_TE_adh})-(\ref{e_TS_CVE_adh}), shows that the TE, VE, and CVE agree
for all examined values of $P$ and, in particular, converge to the same value
for $P\rightarrow\infty$. The three estimators, however, differ in their
statistical convergence. Unlike for the unconstrained result, where the
CVE is the optimal estimator, for the constrained energy, the optimal estimator is
the VE. This can be clearly seen in Fig. \ref{fig:eckart_EvsP_adhoc} (b) which
shows the RMSEs of the different estimators for
different values of $P$. While the RMSE of the TE increases with $P$, the
RMSEs of the VE and CVE remain approximately constant as a function of $P$,
with the VE having a much smaller statistical error. Assuming that the desired
convergence is achieved for $P=24$, the speedup factor achieved by using the VE
compared to the TE and CVE is approximately $2.9^{2}\approx8$ and
$8.1^{2}\approx60,$ respectively. It is clear from the figure that both the speedup factor and the best estimator depend on $P$ and hence on the temperature. 

Figure \ref{fig:eckart_EvsP_rigorous} shows
the same results, but computed with the estimators (\ref{e_TS_TE_rig}%
)-(\ref{e_TS_CVE_rig}). The statistical errors are very similar, although for the VE
slightly larger than those in Fig. \ref{fig:eckart_EvsP_adhoc}.


\subsection{The $\mathrm{H+H_{2}\rightarrow H_{2}+H}$ reaction}

The temperature dependence of the rate constant of the $\mathrm{H+H_{2}%
\rightarrow H_{2}+H}$ reaction was studied on the
Boothroyd-Keogh-Martin-Peterson (BKMP2) reactive potential energy
surface.\cite{Boothroyd1991,Boothroyd1996,Boothroyd} The classical transition
state of this system has a collinear configuration with equal bond lengths
$d_{\mathrm{H_{a}H_{b}}}=d_{\mathrm{H_{b}H_{c}}}$. A suitable reaction
coordinate is therefore given by the difference of the bond lengths,
\begin{equation}
\xi(\mathbf{r}):=d_{\mathrm{H_{a}H_{b}}}-d_{\mathrm{H_{b}H_{c}}}.
\end{equation}

\begin{figure}[hptb]
\includegraphics[width=1\columnwidth]{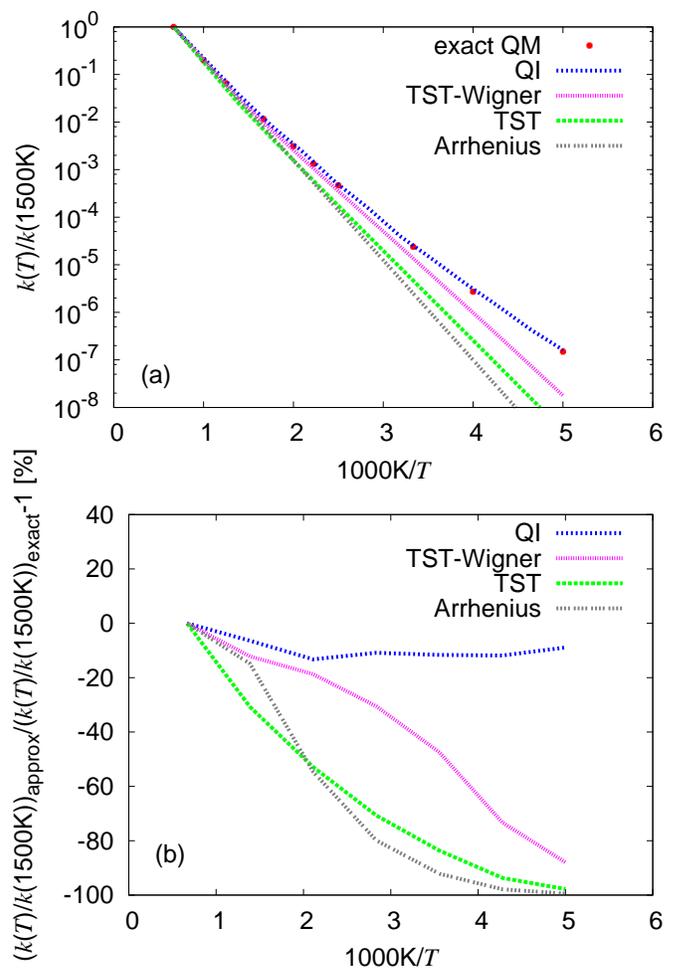}\caption{The H$\,+\mathrm{\,}$
H$_{2}\rightarrow\,$H$_{2}+\,$H reaction. (a) Temperature dependence of the
rate constant. (b) Temperature dependence of the relative error of the ratio
$k(T)/k(1500\operatorname{K})$.}
\label{fig:h3}%
\end{figure}

Figure \ref{fig:h3} (a) shows the temperature dependence of the rate constant
in the range from $200%
\operatorname{K}%
$ to $1500%
\operatorname{K}%
$. The exact QM\ results are from Ref. \onlinecite{Vanicek2005}. At high
temperatures the TST curve is tangent to the exact QM curve, but at low
temperatures, there is a significant discrepancy even for the TST with the
Wigner tunneling correction. On the other hand, the QI approximation agrees
very well with the exact QM result all the way to $200%
\operatorname{K}%
$. The relative error of the rate constant ratio is shown in Fig. \ref{fig:h3}
(b) which confirms that the error of the QI approach is within $13\%$ in the
full temperature range whereas all other approximations have huge errors
already for temperatures as high as $500%
\operatorname{K}%
$.

In case of the $\mathrm{H+H_{2}\rightarrow H_{2}+H}$ reaction, the VE had to
be corrected for both $E_r$ and $E^{\ddag}$ calculations. The reason is that the virial theorem only holds for bound
systems. The transition state of the $\text{H}_{3}$ system can translate
freely as a whole and the three translational degrees of freedom yield a
correction of $D/(2\beta)=3/(2\beta)$ to the VE. In the reactant region, both
the $\text{H}$ atom and $\text{H}_{2}$ molecule can move freely and the six
translational degrees of freedom give a correction of $6/(2\beta)$ to the VE.

\begin{figure}[hptb]
\includegraphics[width=1\columnwidth]{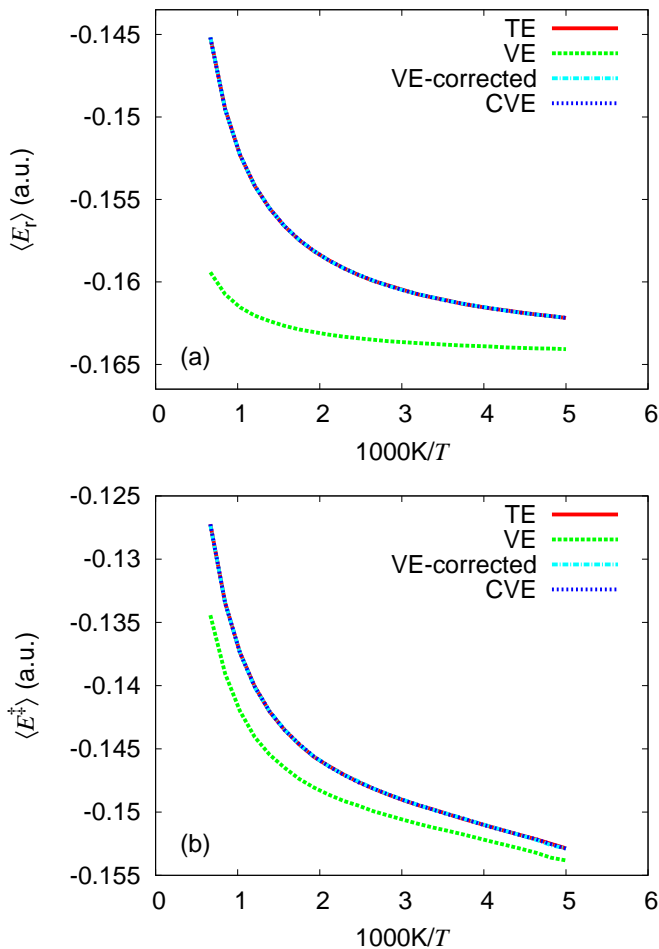}\caption{The H$\,+\mathrm{\,}$
H$_{2}\rightarrow\,$H$_{2}+\,$H reaction. Temperature dependence of the reactant (a) and transition state (b) energies.}
\label{fig:h3_ti}%
\end{figure}

The temperature dependence of the reactant and transition state energies is shown in Fig.~\ref{fig:h3_ti}. While both curves are quite smooth, small discretization errors in the integrals can have large effects on the exponentiated result. By an error analysis described in Sec. III, we found that the Simpson method was sufficient for the TI over the whole temperature range.

\begin{figure}[hptb]
\includegraphics[width=1\columnwidth]{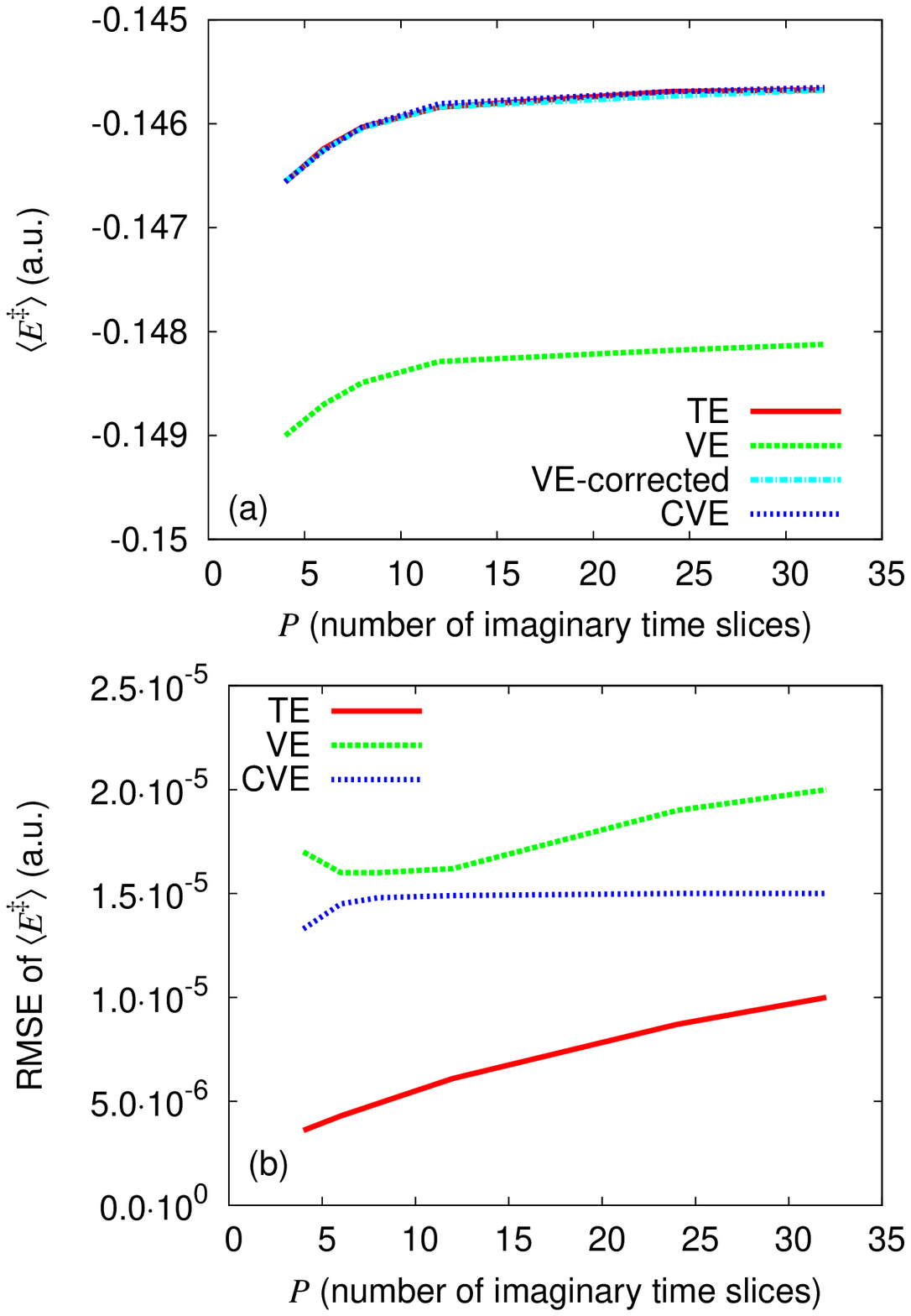}\caption{Dependence of the
transition state energy (a) and of the statistical RMSE of the transition state energy (b)
on the number of imaginary time slices for the H$\,+\mathrm{\,}$%
H$_{2}\rightarrow\,$H$_{2}+\,$H reaction at $T=515.15\operatorname{K}$. The
constraint potential is independent of temperature and estimators
(\ref{e_TS_TE_adh})-(\ref{e_TS_CVE_adh}) are used.}
\label{fig:h3_EvsP_adhoc}%
\end{figure}

\begin{figure}[hptb]
\includegraphics[width=1\columnwidth]{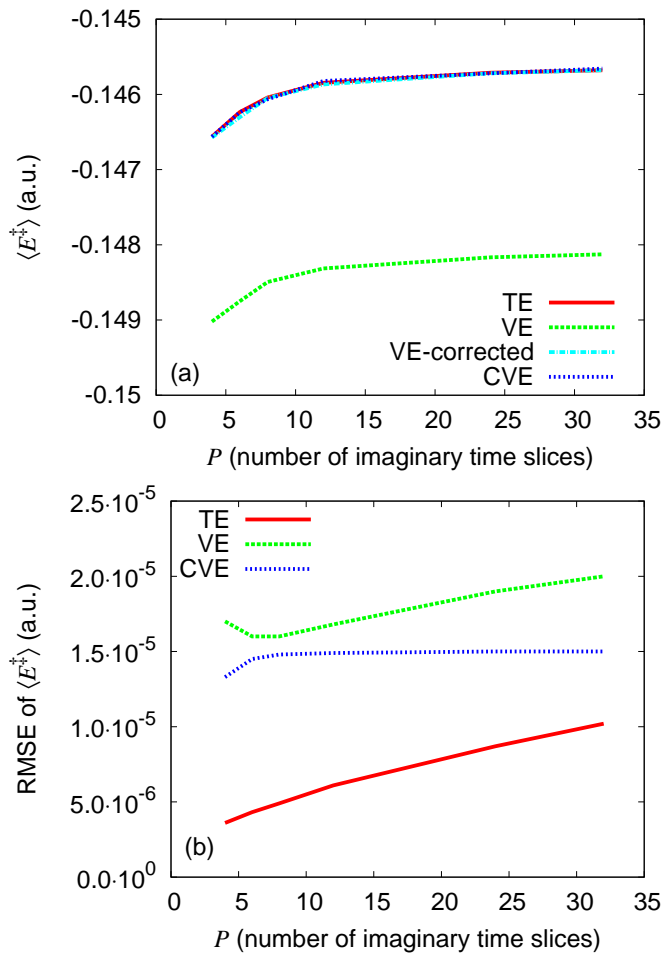}\caption{Dependence of the transition
state energy (a) and of the statistical RMSE of the transition state energy (b) on the
number of imaginary time slices for the H$\,+\mathrm{\,}$H$_{2}\rightarrow
\,$H$_{2}+\,$H reaction at $T=515.15\operatorname{K}$. The constraint
potential is proportional to temperature and estimators (\ref{e_TS_TE_adh})-(\ref{e_TS_CVE_adh})
 are used.}
\label{fig:h3_EvsP_vconstr_propto_T}%
\end{figure}

Figures \ref{fig:h3_EvsP_adhoc} and \ref{fig:h3_EvsP_vconstr_propto_T} show
how the constraint energy $E^{\ddag}$ and the RMSE of $E^{\ddag}$ depend on
$P$ for $T=515.15\operatorname{K}$. Panel (a) of Fig. \ref{fig:h3_EvsP_adhoc},
which uses the estimators (\ref{e_TS_TE_adh})-(\ref{e_TS_CVE_adh}) and 
a constraint potential independent of temperature,
shows again that the TE, corrected VE, and CVE give approximately the
same results for all values of $P$ and, within a statistical error, converge
to the same limiting value for $P\rightarrow\infty$. Panel (b) of Fig.
\ref{fig:h3_EvsP_adhoc} shows that while the statistical error of the CVE is
approximately constant as a function of $P$, the RMSEs of the TE\ and VE grow
with $P$. However, in this case, the results are quite well converged for $P=32$ and
at this point the RMSE of the CVE is still larger than the RMSE of the TE, although 
it is already smaller than the RMSE of the VE.
While for lower temperatures where larger values of $P$ are needed, CVE would
eventually become the optimal estimator, it is not so for
$T=515.15\operatorname{K}$. The growth of the RMSE of the VE with $P$ is due
to the fact that unlike for the Eckart barrier, the transition state of the H$_{3}$ system can
move freely as a whole.

Finally, Fig. \ref{fig:h3_EvsP_vconstr_propto_T} shows analogous results, still 
using estimators (\ref{e_TS_TE_adh})-(\ref{e_TS_CVE_adh}), but
obtained with a constraint potential (\ref{v_constr_propto_T}) proportional to
$T$ (chosen such that the two
types of constraints coincide for $T=515.15%
\operatorname{K}%
$). As expected, the statistical errors of the TE, VE, and CVE are similar to those in Fig.
\ref{fig:h3_EvsP_adhoc} [obtained with the constraint potential
(\ref{v_constr}) independent of $T$].

\section{Conclusions}

A general method for the direct evaluation of the temperature dependence of
the quantum rate constant was presented.
The main advantage of this method is the increased efficiency: Evaluating the
temperature dependence directly, without computing the rate constant at any
given temperature, allows us to avoid a tedious umbrella sampling procedure. 

Besides efficiency, the direct calculation of the temperature
dependence of the rate constant can also improve the accuracy:  
Our ratios $k_{\text{QI}}(T)/k_{\text{QI}}(1500\operatorname{K})$ for both the Eckart barrier and the $\mathrm{H+H_{2}\rightarrow
H_{2}+H}$ reaction have somewhat smaller relative errors than the errors obtained for the absolute QI rate constants in previous studies of these systems.\cite{Miller2003,Yamamoto2004}  
The smaller relative error in the ratio of rate constants is due to a favorable cancellation of various systematic errors, such as the systematic error of about $25\%$ of the QI model at high temperatures (that can also be removed by an ad hoc correction of $\Delta H$)\cite{Miller2003} and small recrossing effects in the $\mathrm{H+H_{2}\rightarrow H_{2}+H}$ reaction, also at high temperatures.

It is noteworthy that for both reactions, the RMSEs of transition state energies
depend on the strength of the constraint. Weakening the constraint facilitates sampling of the configuration space and the error of the CVE
decreases, approaching the well-known unconstrained situation where the CVE is typically
the optimal estimator. However, at the same time the constraint must be strong enough to exert the constraining effect and describe the situation near the transition state properly. As a result, ``ranking'' of the estimators is not universal but can change with the potential used as a constraint and is
in general different from the ranking for unconstrained simulations.

The dependence of the error of the VE on $P$ is best understood in terms of the ring polymer 
interpretation\cite{Chandler1981} of the discretized PI: The quantum thermodynamics of the 
original system can be interpreted as the classical thermodynamics of the ring polymer. 
The constrained PI simulation for the one-dimensional Eckart barrier 
is completely bound, resulting in the RMSE independent of $P$. In the full-dimensional hydrogen 
exchange reaction, on the other hand, even the constrained simulation allows the system 
to move as a whole. This is exactly 
where the VE is known to have a RMSE increasing with $P$.

The CVE estimator is usually the optimal estimator in unconstrained systems with some translational 
(i.e., free-particle)  degrees of freedom. In a system in which only two slices 
are bound (in our case, slices $P/2$ and $P$) , the symmetry between different slices is 
lost and so is, to some extent, 
the advantage of subtracting the centroid. This explains why the RMSE of the VE can 
sometimes be smaller than the RMSE of the CVE for all values of $P$, which we observed 
in Figs. \ref{fig:eckart_EvsP_adhoc} and \ref{fig:eckart_EvsP_rigorous}.

To sum up, while in generic systems at very low temperatures, the CVE is expected to be 
the optimal estimator for energy, at finite temperatures in constrained simulations, the VE or 
even the TE can have the smallest RMSE. The results obtained in this paper can serve as a guide
for choosing the best estimator for a given system. However, since the additional cost of 
evaluating all three estimators is negligible in comparison to the cost of the PIMC random 
walk or PIMD simulation, we recommend computing all three estimators, evaluating their RMSEs, 
and using the one with the smallest RMSE in a given situation.


\section{Acknowledgments}

This research was supported by the Swiss NSF (Grant No. $200021\_124936/1$)
and by the EPFL. Authors thank Tom\'{a}\v{s} Zimmermann for helpful discussions.

\bibliographystyle{apsrev}

\end{document}